%% file: doublylocal.tex
\documentclass{jfm}

\usepackage{graphicx}
\usepackage{natbib}
\usepackage{dcolumn}
\usepackage{upmath}

\input defs

\graphicspath{{./figs-png-small/}}

\title[A doubly-localized equilibrium of plane Couette flow]
{A doubly-localized equilibrium solution of plane Couette flow}

\author[
E.\ Brand
    and
J.F.\ Gibson
]
{
E.\ns  B\ls R\ls A\ls N\ls D
\and
J.\ns F.\ns G\ls I\ls B\ls S\ls O\ls N
}

\affiliation{
Dept.\ of Mathematics and Statistics, University of New Hampshire,
Durham, NH 03824, USA
}

\date{\today}
\begin{document}
\maketitle

\begin{abstract}
We present an equilibrium solution of plane Couette flow that is
exponentially localized in both the spanwise and streamwise
directions. The solution is similar in size and structure to
previously computed turbulent spots and localized, chaotically
wandering edge states of plane Couette flow. A linear analysis of
dominant terms in the Navier-Stokes equations shows how the
exponential decay rate and the wall-normal overhang profile of the
streamwise tails are governed by the Reynolds number and the 
dominant spanwise wavenumber. Perturbations of the solution 
along its leading eigenfunctions cause rapid disruption of the 
interior roll-streak structure and formation of a turbulent spot, 
whose growth or decay depends on the Reynolds number and the 
choice of perturbation. 
\end{abstract}

\input intro

\input numerics

\input solutions

\input conclusions

\vspace*{0.6cm}
\noindent{\bf Acknowledgments.}
The authors thank Tobias Schneider, Greg Chini, and Bruno Eckhardt for 
helpful discussions, Hecke Schrobsdorff and Tobias Kreilos for their work on 
parallellizing {\tt channelflow}, and the Max Planck Institute for Dynamics 
and Self-Organization for computer time. 

\bibliographystyle{jfm}
\bibliography{doublylocal}

\end{document}

%% file: defs.tex
\ifCUPmtlplainloaded \else
  \checkfont{eurm10}
  \iffontfound
    \IfFileExists{upmath.sty}
      {\typeout{^^JFound AMS Euler Roman fonts on the system,
                   using the 'upmath' package.^^J}%
       \usepackage{upmath}}
      {\typeout{^^JFound AMS Euler Roman fonts on the system, but you
                   dont seem to have the}%
       \typeout{'upmath' package installed. JFM.cls can take advantage
                 of these fonts,^^Jif you use 'upmath' package.^^J}%
       \providecommand\upi{\pi}%
      }
  \else
    \providecommand\upi{\pi}%
  \fi
\fi

\ifCUPmtlplainloaded \else
  \checkfont{msam10}
  \iffontfound
    \IfFileExists{amssymb.sty}
      {\typeout{^^JFound AMS Symbol fonts on the system, using the
                'amssymb' package.^^J}%
       \usepackage{amssymb}%
         \let\leq=\leqslant
         \let\geq=\geqslant
      }{}
  \fi
\fi

\ifCUPmtlplainloaded \else
  \IfFileExists{amsbsy.sty}
    {\typeout{^^JFound the 'amsbsy' package on the system, using it.^^J}%
     \usepackage{amsbsy}}
    {\providecommand\boldsymbol[1]{\mbox{\boldmath $##1$}}}
\fi

\newif\ifdraft
\drafttrue     

\usepackage{color,graphicx}
\usepackage{amsmath,amsfonts,amssymb}
\usepackage{natbib}
\usepackage{subfigure}

\newcommand{\revision} [1] {{#1}}

\newcommand\ui{{\text i}}
\newcommand{\bx}{\ensuremath{{\bf x}}}
\newcommand{\grad}{\ensuremath{{\bf \nabla}}}

\newcommand{\bu}{\ensuremath{{\bf u}}}
\newcommand{\bv}{\ensuremath{{\bf v}}}
\newcommand{\hu}{\ensuremath{\hat{u}}}
\newcommand{\bff}{\ensuremath{{\bf f}}}
\newcommand{\be}{\ensuremath{{\bf e}}}

\newcommand{\butot}{\ensuremath{{\bf u}_{\text{tot}}}}
\newcommand{\Rey}{\mbox{\textit{Re}}}

\newcommand{\refeq}  [1] {(\ref{#1})}

\newcommand{\reffig} [1] {figure~\ref{#1}}

\newcommand{\refFig} [1] {Figure~\ref{#1}}

\newcommand{\refTab} [1] {Table~\ref{#1}}

\newcommand{\refsec}[1] {\S\,\ref{#1}}

\newcommand{\sz}{\ensuremath{\sigma_{z}}}
\newcommand{\sxy}{\ensuremath{\sigma_{xy}}}

\newcommand{\sxyz}{\ensuremath{\sigma_{xyz}}}
\newcommand{\Ai}{\operatorname{Ai}}
\newcommand{\Bi}{\operatorname{Bi}}


%% file: intro.tex
\section{Introduction}

Since the work of \citet{NagataJFM90}
a large number of unstable nonlinear equilibrium, traveling-wave, and 
periodic-orbit solutions of the Navier-Stokes equations have been 
computed for a variety of canonical flows including pipe, channel,
plane Couette, and square-duct flow. These invariant solutions demonstrate the 
feasibility and fruitfulness of treating well-resolved direct numerical 
simulations as very-high-dimensional dynamical systems, and they
capture, in a precise and elemental form, a number of important coherent flow 
structures and dynamical processes. Linear stability analysis shows that 
these solutions have relatively few unstable modes, and that the solutions
and their low-dimensional unstable manifolds impose structure on the 
dynamics of moderately turbulent flows. See \citet{KawaharaARFM12} for a recent 
review of this work.
Most of this work has been done in the context of canonical flows in 
small computational domains with periodic boundary conditions,
resulting in spatially periodic solutions that lie within dynamically 
invariant periodic subspaces of the same flows on infinite domains. 
While small periodic `minimal flow units' are useful microcosms for
studying turbulence, turbulence in extended domains generally involves
large numbers of interacting flow structures, whose dynamic coupling 
presumably decreases with their separation. Additionally the transition 
to turbulence in extended domains occurs through the growth of turbulent 
spots or puffs, consisting of localized patches of unsteady, complex flow 
within a background of laminar flow \citep{WygnanskiJFM73,
TillmarkJFM92,BarkleyPRL05,PhilipPRE11}. 

These considerations motivate the search for spatially-localized invariant 
solutions of flows in extended domains. \citet{SchneiderJFM10} found the first known 
localized solutions, a pair of spanwise-localized, streamwise-periodic 
equilibrium and traveling-wave solutions of plane Couette flow, further
investigated in \citet{SchneiderPRL10}. \citet{AvilaPRL13} found a 
streamwise-localized relative periodic orbit of pipe flow that closely resembles 
the transient turbulent puffs of \citet{HofNature06}. \citet{DeguchiJFM13} 
and \citet{GibsonJFM14} independently found spanwise-localized 
forms of the periodic EQ7/HVS solution of \citet{ItanoPRL09,GibsonJFM09}. 
\citet{GibsonJFM14} also presented a number of spanwise-localized and 
wall-normal-localized traveling waves of channel flow. \cite{KhapkoJFM13} found
spanwise-localized relative periodic orbits of the asymptotic suction boundary
layer, and \cite{ZammertARX14} found a spanwise- and wall-normal-localized 
periodic orbit of plane Poiseuille flow.

This paper presents a span- and streamwise-localized equilibrium solution of 
plane Couette flow, the first known invariant solution of the Navier-Stokes 
equations localized in two homogeneous directions. The numerical procedure by 
which the doubly-localized solution was found is outlined in \refsec{s:numerics}. 
Properties of the solution are presented in \refsec{s:solutions}, including its 
exponential localization, its global quadrupolar flow, the geometrical structure 
of its rolls and streaks, its wall-normal overhang profile, and the role of its 
instabilities in the transition to turbulence.

%% file: numerics.tex
\section{Computation of doubly-localized solutions}
\label{s:numerics}

The mathematical formulation and numerical methods are presented in 
detail in \citet{GibsonJFM14} (GB14); here we present
a brief outline. The Reynolds number $\Rey$ for plane Couette flow 
is defined in terms of half the relative wall speed, the channel half-height, and 
the kinematic viscosity, so that the walls at $y = \pm 1$ have velocity $\pm 1$ 
and the laminar flow solution is given by $y \, \be_x$. The total velocity is
expressed as a sum of the laminar flow and the deviation from laminar, 
$\butot = y \, \be_x + \bu$, and henceforth we refer to the deviation
$\bu = [u,v,w]$ as velocity. With these assumptions $\bu$ has zero Dirichlet 
boundary conditions at the walls, and the nondimensionalized Navier-Stokes 
equations take the form 
\begin{equation} 
\frac{\partial \bu}{\partial t} + y \frac{\partial \bu}{\partial x} + v \: \be_x  
 + \bu \cdot \grad \bu = -\nabla p  + \frac {1} {\Rey} \nabla^2 \bu, \quad \nabla \cdot \bu  = 0.
\label{eq:NSE_PCF}
\end{equation}
The nondimensionalized computational domain is 
$[-L_x/2, L_x/2] \times [-1, 1] \times [-L_z/2, L_z/2]$ with periodic 
boundary conditions in the streamwise $x$ and spanwise $z$ directions. 
Discretization is performed with standard Fourier-Chebyshev spectral 
methods in space, 3rd-order semi-implicit finite differencing in time, and 2/3-style 
dealiasing. The computational domain and spatial discretization are specified
in terms of $L_x \times L_z$ and the collocation grid $N_x \times N_y \times N_z$. 
Equilibria are computed as solutions of $\bff^T(\bu) - \bu = 0$, where $\bff^T$ is the 
time integration of \refeq{eq:NSE_PCF} for a fixed time $T$, and 
the discretized equations are solved with a Newton-Krylov-hookstep search algorithm
\citep{ViswanathJFM07,ViswanathPTRSA09}. \revision{The choice for the $T$ is
determined by a practical balance in the computational solution of the Newton-step
equation: too small a value of $T$ results in weak viscous damping and slow convergence of 
the iterative GMRES algorithm, but too large $T$ reduces the distance $\| \delta \bu \|$ 
over which the linearization $f^T(\bu + \delta \bu) \approx f^T(\bu) + Df^T \delta \bu$ 
is accurate. We have found that $T=O(10)$ is a good balance for a wide variety 
of flow conditions and Reynolds numbers.}
The software \revision{and the numerical data for the doubly-localized solution} are available at 
{\tt www.channelflow.org} \citep{GibsonJFM08,chflow}.

Initial guesses for the doubly-localized solutions were produced by applying 
streamwise windowing to the spanwise-localized forms of EQ7 solution from GB14,
or two-dimensional windowing to the doubly-periodic EQ7 solution from 
\citet{GibsonJFM09}. We used the same $\tanh$-based windowing function as in 
GB14 equation (2.4), replacing $z$ with $x$ for a streamwise windowing function 
\begin{equation}
W(x) = \frac{1}{4} \; \left(1 + \tanh \left(\frac{6(a-x)}{b} + 3\right)\right) 
                      \left(1 + \tanh \left(\frac{6(a+x)}{b} + 3\right)\right).
\label{eq:window}
\end{equation}
As noted in GB14, $W(x)$ is even, smooth, monotonic in $|x|$, and close to unity 
for the core region $|x| < a$, transitions smoothly to nearly zero over 
$a < |x| < a + b$, and approaches zero exponentially as $|x| \rightarrow \infty$. 
The nonzero divergence of windowed velocity fields $W(x) \bu(x,y,z)$ is fixed 
by revising the wall-normal $v$ component to satisfy incompressibility. To create 
doubly-localized initial guesses from  doubly-periodic solutions, we applied the 
two-dimensional windowing function $W(x)W(z)$ with different length scales for 
the core and transition regions in the streamwise and spanwise directions. 

It was considerably more difficult to find doubly-localized solutions from
windowed initial guesses than it was to find the spanwise-localized solutions 
of GB14. Not only do the doubly-localized solutions require doubly-extended 
domains, the decay rate of their tails is slower than for the spanwise-localized 
solutions (see \refsec{s:xtails}), and thus the computational domains must be 
larger in both $x$ and $z$. Search results were sensitive both to
the wavelengths of the underlying periodic or spanwise localized solution and
to the choice of windowing parameters. The search landscape for doubly-localized 
solutions is also vastly more complicated and more sensitive to spatial 
discretization, with many nonzero local minima and many spurious solutions for 
under-resolved discretizations. Lastly, doubly-localized initial guesses tended 
to converge onto the trivial solution $\bu = 0$ (laminar flow), with the search quickly 
settling onto streaky flow with very little streamwise variation, and then reducing 
the magnitude of the streaks to zero. Such streaks are dynamical transients that 
decay to laminar flow under time evolution, but their decay is slow enough that 
they nearly satisfy the search equation $\bff^T(\bu) - \bu = 0$ for small $T$, 
thus attracting nearby guesses to a search path that ultimately leads to $\bu = 0$. 

To prevent the search algorithm from being fooled by such transients, we modified 
the search equation to $(\bff^T(\bu) - \bu)/(\|\bu\|_{3d} - c) = 0$, where 
$\| \bu \|_{3d}$ is the \revision{energy norm (see \refsec{s:stability})} of the 
streamwise-varying portion 
of $\bu$ and $c$ is a parameter set to some fraction of the value of $\| \bu \|_{3d}$ 
for the initial guess. Our choices for underlying periodicity and windowing parameters 
were determined by trial and error, guided by the length scales that approximate 
solutions took on during the search. To mitigate computational costs, we performed 
trial-and-error calculations in relatively small domains with poor localization 
($80 \times 20$ and $O(10^{-1})$ tails at the perimeter) and then extended solutions 
from successful searches to larger domains where localization is more pronounced 
($200 \times 200$ with $O(10^{-3})$ tails), using either continuation in $L_x,L_z$ 
or simply doubling the computational domain and reapplying windowing. For doubly-localized 
solutions we found that solutions were reliably robust to changes in discretization 
when spectral coefficients were retained to $O(10^{-7})$ in $x,z$ and $O(10^{-10})$ in $y$. 
The solution presented in the following section is the most robust of several 
we found, in that it converges quickly at higher spatial resolutions and continues 
smoothly and easily in $L_x,L_z$ and $\Rey$. 
This solution was found by applying two-dimensional 
windowing to the doubly-periodic EQ7 solution, refinement to an exact solution 
by Newton-Krylov-hookstep, and extension to large domains by repeated doubling, 
windowing, and refinement.

%% file: solutions.tex
\section{Properties of the doubly-localized solution}
\label{s:solutions}

\subsection{Global flow}
\label{s:global}

\begin{figure}
{\footnotesize (a)}  \hspace{-2mm} \includegraphics[width=0.45\textwidth]{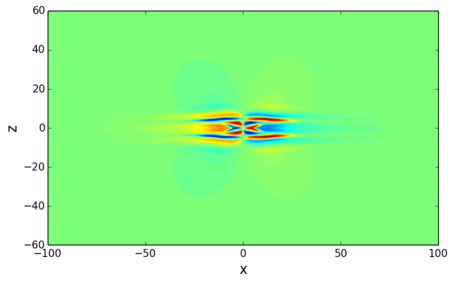} 
{\footnotesize (b)}  \hspace{-2mm} \includegraphics[width=0.45\textwidth]{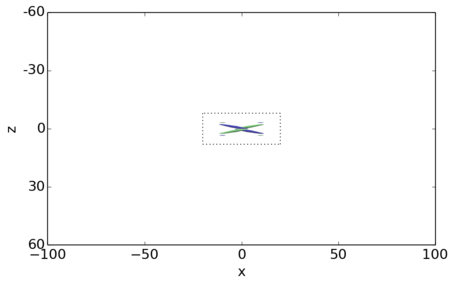} \\
{\footnotesize (c)}  \hspace{-2mm} \includegraphics[width=0.45\textwidth]{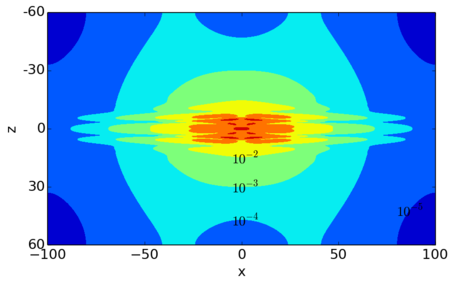} 
{\footnotesize (d)}  \hspace{-2mm} \includegraphics[width=0.45\textwidth]{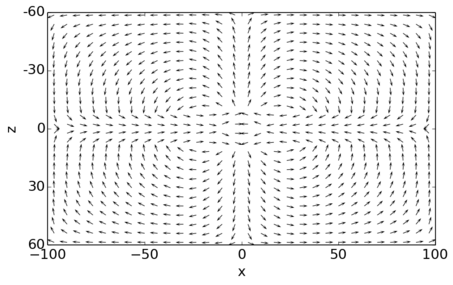} 
\caption{{\bf A doubly-localized equilibrium solution of plane Couette flow} at $\Rey=240$
in a $200 \times 120$ computational domain.
(a) Streamwise velocity $u$ in the $y=0$ midplane. The blue-red color axis spans 
$u \in [-0.5, 0.5]$. 
(b) Isosurfaces of swirling strength at $s=\pm0.2$ in green/blue, indicating swirling 
with clockwise/counterclockwise orientation with respect to the positive $x$ axis. 
The dotted $40 \times 16$ subdomain is shown in detail in \reffig{f:swirling_detail}.
(c) Exponential localization, indicated by contours of $y$-integrated energy (see text). 
Contour levels are set at $10^{-n}$ for $n=0,1,\ldots,5$; the boundaries for $10^{-2}$ 
through $10^{-5}$ are labeled.
(d) Quadrupolar $y$-averaged global flow, shown by a vector plot of 
$(\bar{u},\bar{w})/\sqrt{\bar{u}^2 + \bar{w}^2}$, where $\bar{u},\bar{w}$ are the $y$ 
averages of $u,w$. Note that the \revision{vector} spacing in this plot is too coarse to resolve
rapid variations in the region surrounding the origin.
}
\label{f:xzplane_full}
\end{figure}

\refFig{f:xzplane_full} shows a doubly-localized equilibrium solution of plane Couette 
flow in a $200 \times 120$ computational domain at $\Rey=240$, discretized with 
$720 \times 49 \times 1024$ gridpoints. 
\revision{The solution has the symmetry group $\{e, \sxy, \sz, \sxyz\}$ 
where
\begin{align}
\sxy &: [u,v,w](x,y,z) \rightarrow [-u,-v,w](-x,-y,z), \\
\sz  &: [u,v,w](x,y,z) \rightarrow [u,v,-w](x,y,-z), \nonumber
\end{align}
$\sxyz = \sxy \sz$ and $e$ is the identity. We use standard angle-bracket 
notation from group theory to specify groups in terms of their generators,
e.g. $\langle \sxy, \sz \rangle = \{e, \sxy, \sz, \sxyz\}$. The 
doubly-localized solution acquires $\langle \sxy, \sz \rangle$ symmetry from 
the windowing breaking the symmetries of EQ7 that involve $x$ and $z$ translation, 
in the same manner as EQ7-2 of GB14.} The streamwise 
velocity in the $y=0$ midplane, shown in \refFig{f:xzplane_full}(a), is roughly 
comparable to the dynamically wandering doubly-localized edge state at $\Rey=400$ 
shown in figure 5 of \citet{SchneiderJFM10}. Both display patterns of wavy streaks 
that are  spanwise narrow and streamwise elongated, and the significant non-laminar 
structure in both is confined to a roughly $100 \times 20$ subdomain of the flow.
\refFig{f:xzplane_full}(c) shows that the solution is exponentially localized in 
both span- and streamwise directions, via contours of the $y$-integrated energy 
$e(x,z) = 1/2 \int_{-1}^{1} \bu \cdot \bu \, dy$.
The fingers that extend along the $x$ axis are due to small-wavelength,
exponentially decaying streaks of streamwise velocity (see \refsec{s:xtails}). 
The deviation from elliptical contours near the edges \revision{of the 
computational domain is an artifact of the periodicity of the domain, which,
together with the solution symmetries, induces even symmetry of $e$ about 
$x=\pm L_x/2$ and $z=\pm L_z/2$.}
\revision{In larger computational domains we have observed elliptical contours and 
exponential decay over four orders of magnitude, with comparable decay rates 
in $x$ and $z$.}

\refFig{f:xzplane_full}(d) shows the direction of the $y$-averaged flow $(\bar{u}, \bar{w})$
by a vector plot of $(\bar{u},\bar{w})/\sqrt{\bar{u}^2 + \bar{w}^2}$. Note the quadrupolar
character of the $y$-averaged flow, similar to figure 6 of \citet{SchumacherPRE01} and
figure 3 of \citet{DuguetPRL13}.
The $\bar{u}, \bar{w}$ flow is streamwise inward along $z = 0$ and spanwise 
outward along $x = 0$, with a global circulation in each of the four quadrants. 
The alignment of the $y$-averaged flow with the $x$ and $z$ axes also results 
from symmetry: $\sz$ symmetry requires that $u$ and $w$ are
even and odd in $z$, respectively, about $z=0$, and $\sxy$ symmetry requires 
that $\bar{u}$ and $\bar{w}$ are odd and even in $x$ about $x=0$. Periodicity in $x$ and 
$z$ requires the same symmetries about the \revision{$x=\pm L_x/2$ and 
$z=\pm L_z/2$} edges of the computational domain, so that the $y$-averaged flow 
aligns with these edges as well \revision{(see \citet{GibsonJFM09})}.

\subsection{Internal structure }
\label{s:internal}

\begin{figure}
\begin{tabular}{ccc}
\raisebox{22mm}{{\bf z~}} & \raisebox{-0.7mm}{\includegraphics[width=0.0615\textwidth]{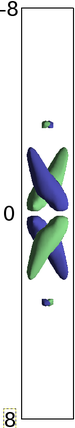}}  & 
    \includegraphics[width=0.85\textwidth]{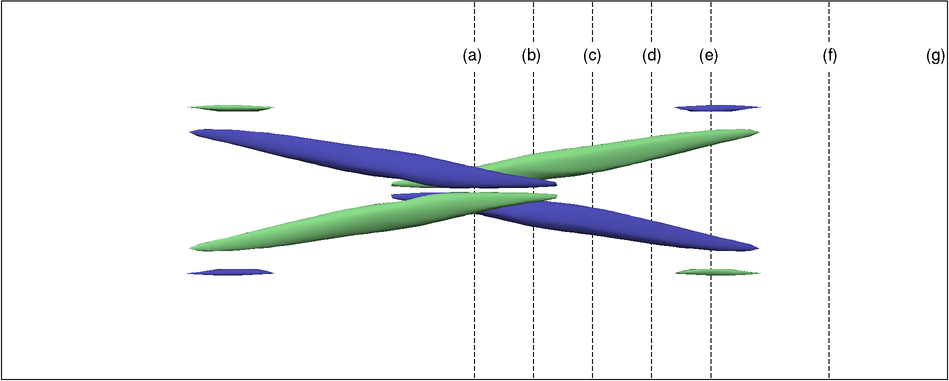} \\
  & \raisebox{-2mm}{\bf ~~y} & \hspace{-1.8mm} \raisebox{-6mm}{\includegraphics[width=0.875\textwidth]{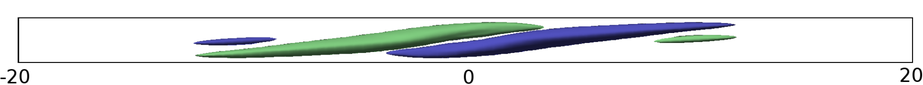}} \\
  &   & \raisebox{-2mm}{\bf x} \\
\end{tabular}
\caption{{\bf Three-dimensional structure of swirling strength.} Isosurfaces of signed swirling 
strength at $s=\pm0.2$ in green/blue. The upper-right subfigure is a blow-up
of the $x \in [-20, 20]$, $z \in [-8, 8]$ subdomain in \reffig{f:xzplane_full}(b). 
Dotted lines show the $x$-positions of the streamwise-normal cross sections depicted 
in \reffig{f:crosssections}, with (a)-(g) at $x = \{0, 2.5, 5, 7.5, 10, 15, 20\}$.
Views of the same structure (left) in $y,z$ and (bottom) in $x,y$. 
}
\label{f:swirling_detail}
\end{figure}

The localized patterns of high- and low-speed streaks shown in \reffig{f:xzplane_full}(a) 
are generated by the highly localized roll structure illustrated in \reffig{f:xzplane_full}(b).
The latter shows isosurfaces of signed swirling strength at $s = \pm 0.2$, about 1/3 of 
its maximum value. \revision{Swirling strength is defined as the magnitude of the
imaginary part of the complex eigenvalues of the velocity gradient tensor $\grad\bu$
\citep{ZhouJFM99}. Signed swirling strength includes a $\pm$ sign indicating the 
orientation of swirling \citep{WuJFM06}, here the sign of the $x$ component of the 
swirling axis when oriented with the right-hand-rule.}
The swirling is highly localized: the magnitude of swirling 
drops by a factor of roughly ten between the X-shaped isosurfaces and the edges of the 
dotted box that marks a $40 \times 16$ subdomain. \refFig{f:swirling_detail} shows a detail of 
the swirling strength in this subdomain with the same plotting conventions as 
\reffig{f:xzplane_full}(b). The three perspective plots show 
an overall X-shaped structure composed of two overlapping $\Lambda$-shaped vortices, whose legs 
swirl in opposite directions and tilt in both the spanwise and wall-normal directions. 
Small, weaker vortices of opposite sign flank the legs near their ends. 

\begin{figure}
\begin{tabular}{c}
{\footnotesize (a)} \includegraphics[width=0.95\textwidth]{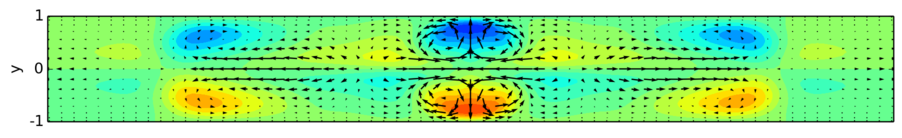} \\
{\footnotesize (b)} \includegraphics[width=0.95\textwidth]{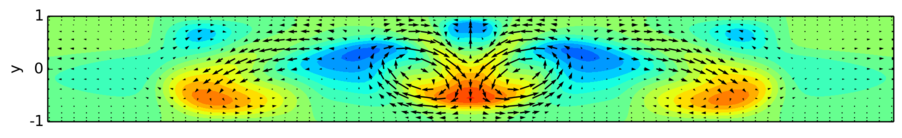} \\
{\footnotesize (c)} \includegraphics[width=0.95\textwidth]{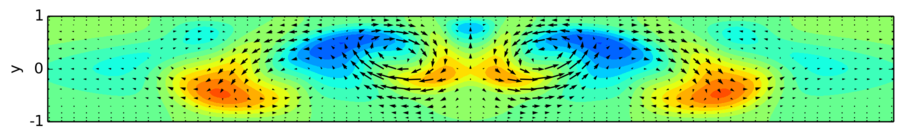} \\
{\footnotesize (d)} \includegraphics[width=0.95\textwidth]{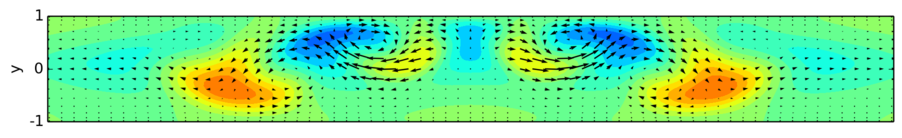} \\
{\footnotesize (e)} \includegraphics[width=0.95\textwidth]{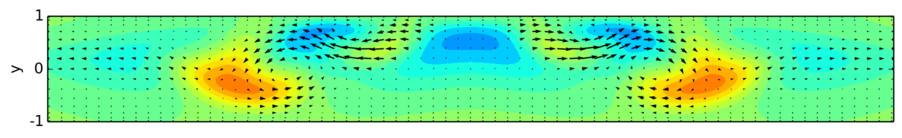} \\
{\footnotesize (f)} \includegraphics[width=0.95\textwidth]{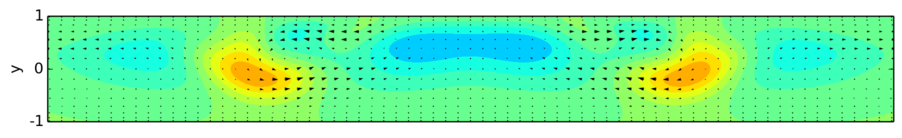} \\
{\footnotesize (g)} \includegraphics[width=0.95\textwidth]{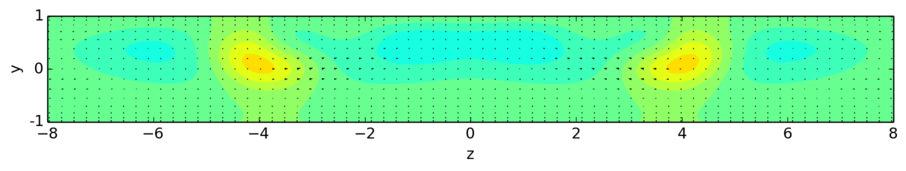} \\
\hspace{8mm}        \includegraphics[width=0.94\textwidth]{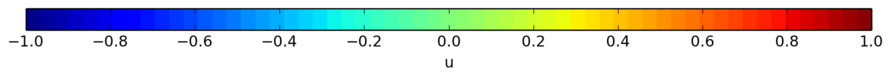} 
\end{tabular}
\caption{{\bf Streamwise-normal cross sections of velocity.} Arrows indicate 
$v,w$ velocity, and color indicates streamwise velocity $u$. The scaling of arrow 
length to $v,w$ magnitude is the same in all graphs. 
(a-g) show $y,z$ planes at $x = \{0, 2.5, 5, 7.5, 10, 15, 20\}$; the positions 
of these planes are marked with dotted lines in \reffig{f:swirling_detail}.
\label{f:crosssections}
} 
\end{figure}

\refFig{f:crosssections} further illustrates the three-dimensional structure of the 
$\Lambda$-shaped vortices and their relation to streamwise streaks. The flow 
in streamwise-normal cross-sections is shown 
at $x$ positions indicated by lines marked (a-g) in \reffig{f:swirling_detail}. 
The $x=0$ plane in (a) shows the $y$-symmetric tips of the two opposed 
$\Lambda$-shaped vortices, concentrated near $z=0$, each drawing midplane fluid 
towards the wall to form streaks. As $x$ increases in (b)-(d), the 
swirling of the legs grows in strength, size, and spacing, 
and moves from the lower wall towards the upper. By (e) the legs 
have reached the upper wall and weakened, but the opposite-signed vortices 
that flank the tips of the legs have grown to their greatest strength. These 
are positioned at about $z = \pm 3$, and they span the distance between the 
walls. By $x=20$ in (g) swirling in both the legs and the flanking vortices
has died out and all that remains are the streamwise streaks.

\subsection{Exponential decay and overhang of the streamwise streaks}
\label{s:xtails}

\begin{figure}
{\footnotesize (a)}  \hspace{-2mm} \includegraphics[width=0.45\textwidth]{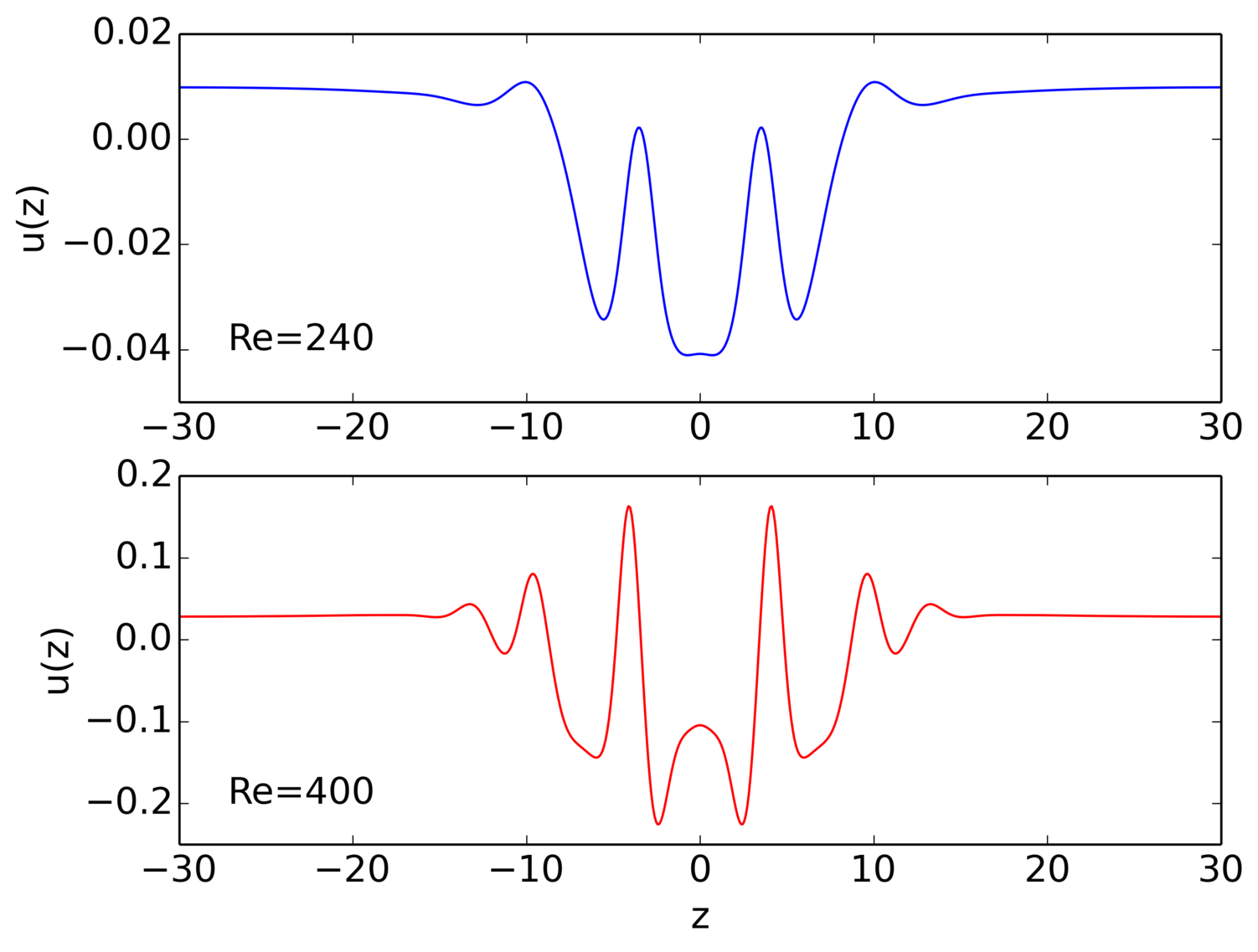} 
{\footnotesize (b)}  \hspace{-2mm} \includegraphics[width=0.45\textwidth]{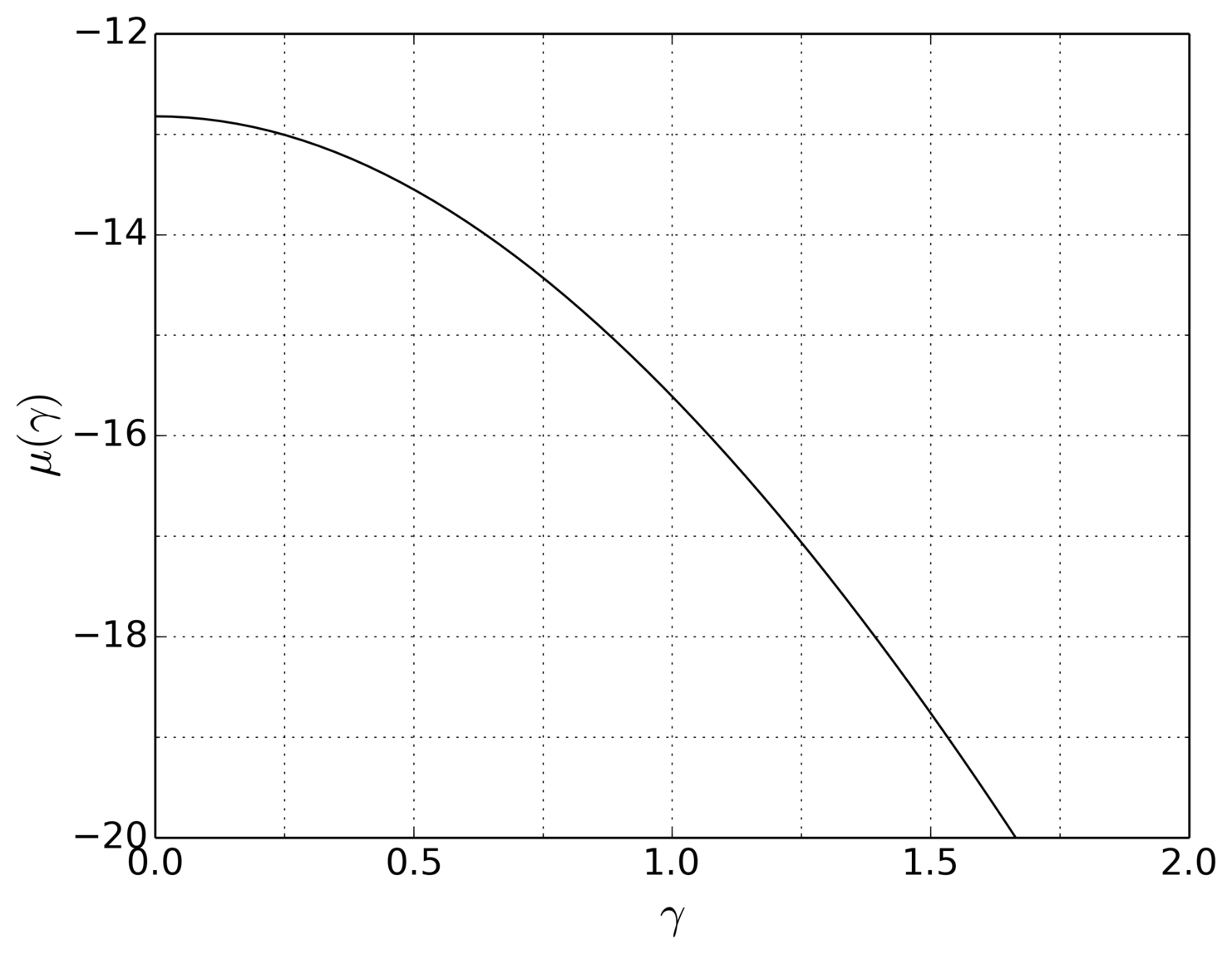} \\
{\footnotesize (c)}  \hspace{-2mm} \includegraphics[width=0.45\textwidth]{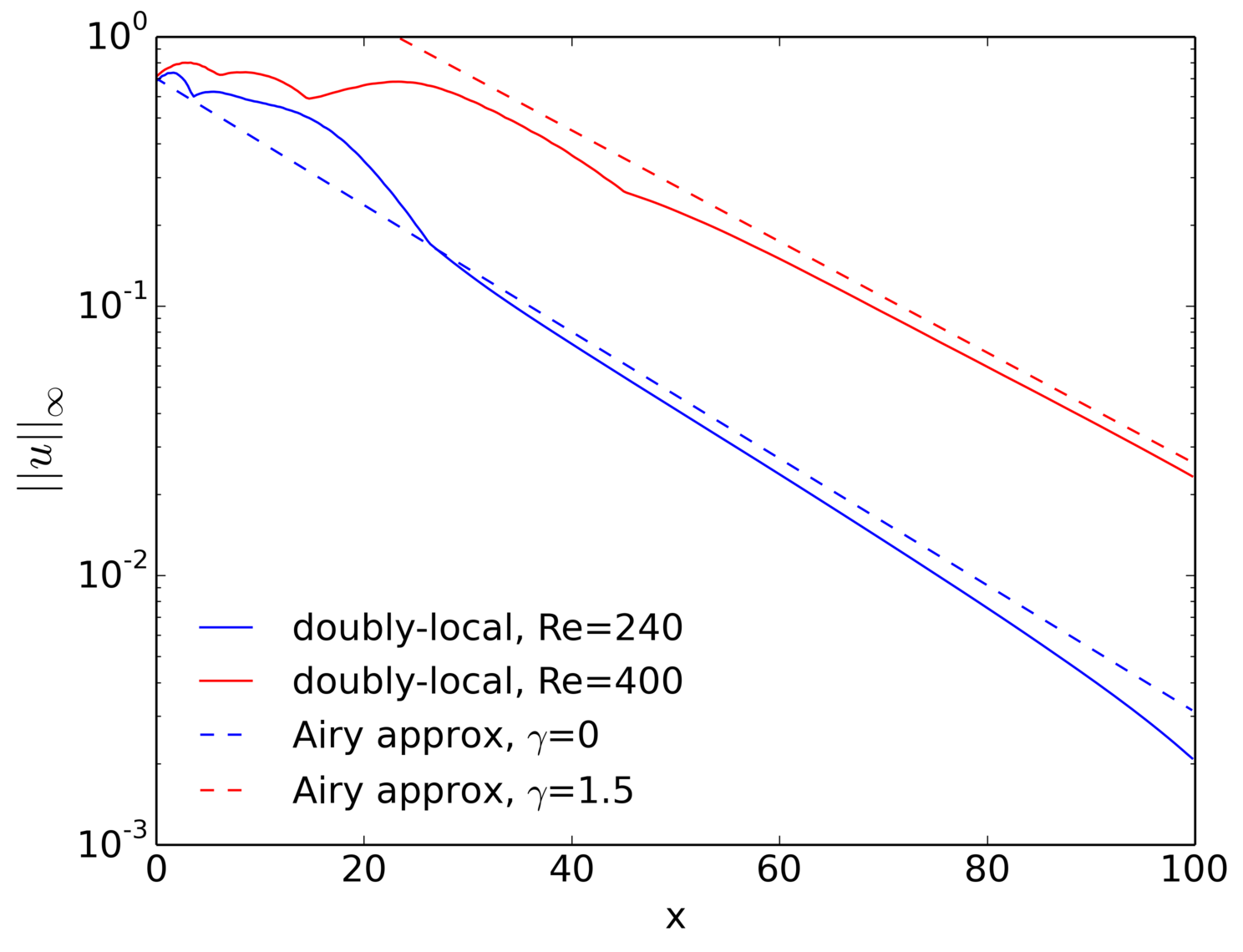} 
{\footnotesize (d)}  \hspace{-2mm} \includegraphics[width=0.45\textwidth]{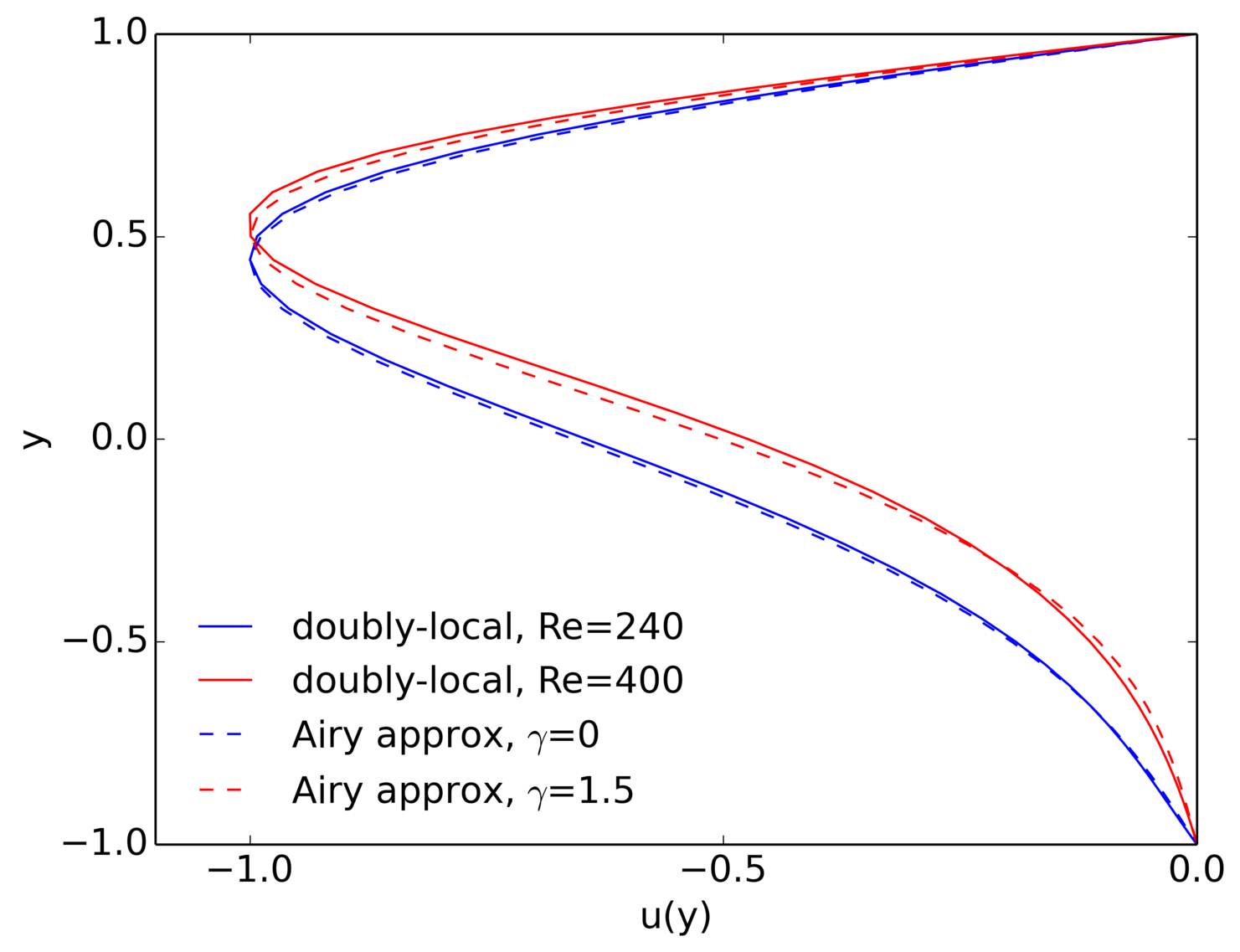}
\caption{{\bf Streamwise decay of streaks.}
(a) The \revision{streamwise velocity $u$ as a function of spanwise 
coordinate $z$ at $x=50$ (within the streamwise tail) and at $y=0.5$ 
(near the peak of the overhang profile).}
(b) Coefficient $\mu$ of exponential streamwise decay rate $\exp(\mu x/ \Rey)$ 
as a function of spanwise wavenumber $\gamma$ for the Airy-function approximation 
of the tails.
(c) Exponential decay of streamwise tails in the doubly-localized solution 
at $\Rey=240$ and $400$ compared to the $\exp(\mu(\gamma)\, x/ \Rey)$  
prediction. \revision{$\|u\|_{\infty}$ denotes the maximum magnitude
of $u$ over $(y,z)$ as a function of $x$.}
(d) Wall-normal overhang profile $u(y)$ at \revision{$x=50$ and $z=0$}, compared 
to predicted $\hat{u}(y)$ (all curves normalized to unit amplitude). 
}
\label{f:xtails}
\end{figure}

The long streamwise tails of the doubly-localized solution are dominated by a
spanwise-localized band of streaky streamwise velocity. Here we provide a 
linear analysis that accounts for the exponential streamwise decay rate of
the streamwise tails and the form of their wall-normal overhang profile, 
effects observed in turbulent spots by \citet{LundbladhJFM91} and \citet{DuguetPRL13}. 
\revision{Direct numerical evaluation} of the magnitudes of different terms in the 
$u$ component of the Navier-Stokes equation \refeq{eq:NSE_PCF} for the doubly-localized
solution shows that the dominant terms in the streamwise tails are
\begin{align}
 y u_x = \Rey^{-1} (u_{yy} + u_{zz})
\label{eq:xtail_pde}
\end{align}
\revision{where subscripts indicate differentiation.}
Although the solution is localized in $z$, a good approximation for the decay 
rate and overhang profile can be obtained by assuming a $z$-periodic solution 
of the form $u(x,y,z) = \hu(y) \exp(\ui \gamma z + \mu x/\Rey)$ with $\gamma$ set 
to match the dominant spanwise wavenumber observed in the streaky tails. 
Substitution of this ansatz into \refeq{eq:xtail_pde} results in the ODE
\begin{align}
\hu''(y) - (\gamma^2 + \mu y) \, \hu(y) = 0,
\label{eq:xtail_ode}
\end{align}
which has solutions 
\begin{align}
\hu(y) = a \Ai(\gamma^2 \mu^{-2/3} + \mu^{1/3} y) + b \Bi(\gamma^2 \mu^{-2/3} + \mu^{1/3} y),
\label{eq:xtail_airysoln}
\end{align}
where $\Ai$ and $\Bi$ are Airy functions. The boundary conditions $\hu(\pm 1) = 0$ 
determine $\mu$ and the relative values of $a$ and $b$ as a function of $\gamma$.
Nontrivial solutions of \refeq{eq:xtail_airysoln} require 
\begin{align}
\Ai( \gamma^2 \mu^{-2/3} + \mu^{1/3})
\Bi( \gamma^2 \mu^{-2/3} - \mu^{1/3}) -
\Ai( \gamma^2 \mu^{-2/3} - \mu^{1/3})
\Bi(  \gamma^2\mu^{-2/3} + \mu^{1/3}) = 0
\label{eq:xtail_eigeneqn}
\end{align}
which we solve numerically for $\mu$ as a function of $\gamma$, choosing the 
negative solution $\mu(\gamma)$ closest to zero in order to find the solution 
with slowest streamwise decay. 

\refFig{f:xtails}(b) shows $\mu$ as a function of $\gamma$. The 
$\gamma \approx 0$ modes have the slowest exponential decay, so 
as $x \rightarrow \infty$ we expect the tails to be dominated by 
the $\gamma=0$ or fundamental $\gamma = 2\upi/L_z$ mode. This behavior 
is evident for the doubly-localized equilibrium at $\Rey=240$. 
\refFig{f:xtails}(a) shows that the streamwise velocity at 
\revision{$x=50$} has a wide slow streak in the region $|z| < 10$ with 
weaker small-scale $z$ variation. The power spectrum of this function 
has \revision{its strongest peak by nearly an order of magnitude}
at the fundamental wavenumber $\gamma = 2\upi/L_z \approx 0.1$. 
The decay rate $\mu(\gamma)$ is nearly constant for small 
$\gamma$, so for $\Rey = 240$ we take $\gamma = 0$ and find 
$\mu(0) \approx -13$. \refFig{f:xtails}(c) shows good agreement 
between the predicted decay $\exp(-13x/240)$ and that observed in 
the doubly-localized solution. \refFig{f:xtails}(d) shows good agreement
between the predicted profile $\hat{u}(y)$ from \refeq{eq:xtail_airysoln} 
and the streamwise velocity profile $u(y)$ of the doubly-localized 
solution \revision{at $x=50$, $z=0$}. 

However, the $1/\Rey$ factor in $\exp(\mu x /\Rey)$ means that 
small $z$-wavelength transients of sufficient magnitude can persist 
and even dominate the tails for $x \ll \Rey$. \revision{\refFig{f:xtails}(a) 
shows that at $x=50$ the small-wavelength $z$ variation for the $\Rey=400$ 
solution is much stronger than for $\Rey = 240$. The power 
spectrum of this $z$ profile has peaks of nearly equal magnitude at 
the fundamental mode $\gamma \approx 0.1$ and at $\gamma \approx 1.5$. 
From what follows we determine that the exponential decay in $\|u\|_{\infty}$ 
for the $z$-localized structure is governed by the slightly less energetic 
$\gamma \approx 1.5$ mode.
Since \revision{$\mu(1.5) \approx -19$} from \reffig{f:xtails}(b),
the decay of this mode is $\exp(-19x/400)$, which is only a factor of two 
smaller than the $\exp(-13x/400)$ decay of the $\gamma=0$ mode over a
length $x \approx 50$. The match of this decay rate in \reffig{f:xtails}(c) 
and the corresponding overhang function in \reffig{f:xtails}(d) to the
doubly-localized solution} confirms that at $\Rey=400$ the tails for 
$x < 80$ are governed by the faster-decaying $\gamma \approx 1.5$ mode. 
For $x = O(\Rey)$ and larger, beyond the limits of the present computational 
domain, we expect the large-$\gamma$ transients to die out leaving the 
tails dominated by more slowly decaying small-$\gamma$ modes. 

\subsection{Stability and the evolution of unstable perturbations}
\label{s:stability}

\begin{table}
\centering
\begin{tabular}{lclclclcl}
& \multicolumn{2}{c}{$\langle \sxy, \sz \rangle$} 
& \multicolumn{2}{c}{$\langle \sxy, -\sz \rangle$} 
& \multicolumn{2}{c}{$\langle -\sxy, \sz \rangle$} 
& \multicolumn{2}{c}{$\langle -\sxy, -\sz \rangle$}\\
$\Rey \quad$ & pos.\ $\lambda_r$ & max.\ $\lambda_r \quad$ & pos.\ $\lambda_r$ & max.\ $\lambda_r \quad$ & pos.\ $\lambda_r$ & max.\ $\lambda_r \quad$ & pos.\ $\lambda_r$ & max.\ $\lambda_r$\\
240 & 2  & 0.0370 & 3  & 0.0414 & 3 & 0.0329 & 8 & 0.1283\\
300 & 4  & 0.0278 & 7  & 0.0314 & 5 & 0.0286 & 14 & 0.1028\\
340 & 6  & 0.0265 & 9  & 0.0635 & 7 & 0.0636 & 15 & 0.0952\\
360 & 8  & 0.0750 & 9  & 0.0751 & 7 & 0.0751 & 15 & 0.0920\\
380 & 8  & 0.0853 & 11  & 0.0854 & 12 & 0.0854 & 19 & 0.0892\\
400 & 17 & 0.0955 & 12 & 0.0958 & 16 & 0.0958 & 23 & 0.0960\\
\end{tabular}
\caption{{\bf Instabilities of the doubly-localized solution.} The number 
of unstable eigenfunctions (positive real part) and the \revision{real part 
of the most unstable eigenvalue} are given for each eigenfunction symmetry group
and a range of Reynolds numbers.
}
\label{t:instabilities}
\end{table}

In minimal flow units, the transition to turbulence is governed by invariant 
`edge state' solutions whose stable manifolds form separatrices between states 
that quickly decay towards laminar flow and states that become turbulent
\revision{\citep{WangPRL07,SchneiderPRE08}}.
Efforts to develop a similar dynamical understanding of transition in extended 
flows have lead to the computation of a number of localized edge states, but to 
date these have either been invariant states localized in a single homogeneous 
direction \citep{SchneiderJFM10,AvilaPRL13,KhapkoJFM13,ZammertARX14} or doubly-localized but chaotically 
wandering states without well-defined stable and unstable manifolds 
\citep{SchneiderJFM10,DuguetPRL12}. The doubly-localized invariant solution 
in this paper thus provides a potential starting point for addressing spatiotemporal 
transition of extended flows in dynamical terms. We focus on a $100 \times 30$ domain, 
large enough to exhibit a range of spatiotemporal behavior \citep{PhilipPRE11}, and 
$230 \leq \Rey \leq 400$, above the $\Rey \approx 228$ saddle-node bifurcation point 
of the \revision{doubly-localized} solution. \refTab{t:instabilities} summarizes the 
properties of the leading unstable eigenfunctions categorized by symmetry group. 
The eigenfunctions $\bv$ of the linearized dynamics about the 
doubly-localized solution are either symmetric \revision{($\bv = \sigma \bv$)} or 
antisymmetric \revision{($\bv = -\sigma \bv$)} for each symmetry of the solution 
and thus have one of four symmetry groups:
$\langle \sxy, \sz \rangle$, 
$\langle \sxy, -\sz \rangle$, 
$\langle -\sxy, \sz \rangle$, or 
$\langle -\sxy, -\sz \rangle$.

\begin{figure}
{\footnotesize (a)}  \hspace{-2mm} \includegraphics[width=0.45\textwidth]{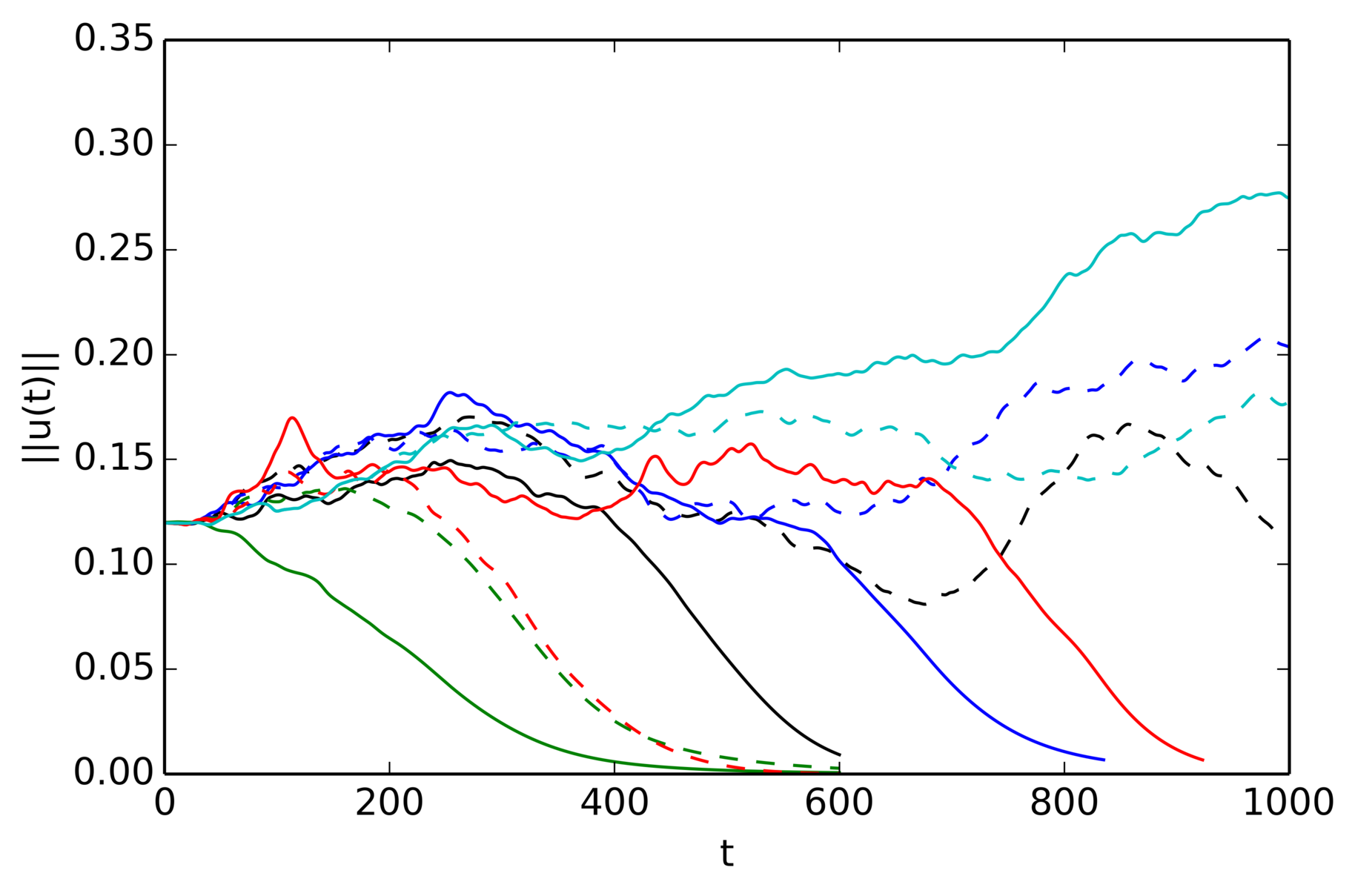} \hspace{2mm}
{\footnotesize (d)}  \hspace{-2mm} \includegraphics[width=0.45\textwidth]{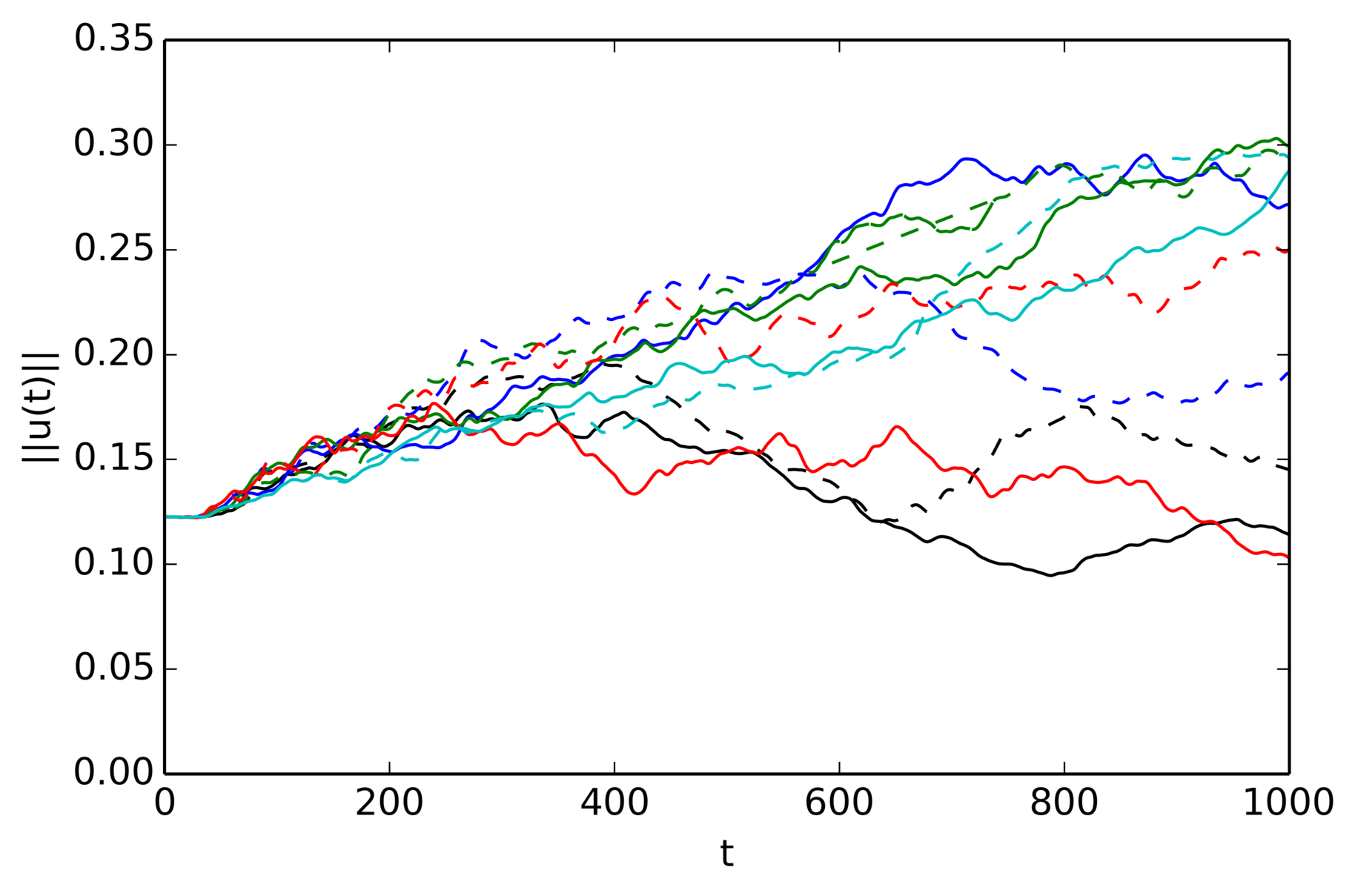} \\
{\footnotesize (b)}  \hspace{-2mm} \includegraphics[width=0.45\textwidth]{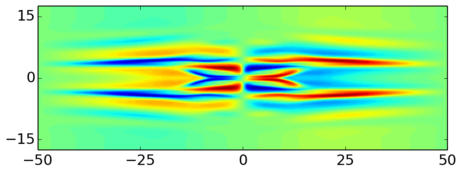} \hspace{2mm}
{\footnotesize (e)}  \hspace{-2mm} \includegraphics[width=0.45\textwidth]{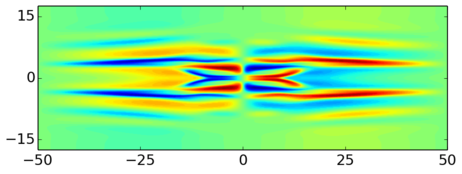} \\
{\footnotesize (c)}  \hspace{-2mm} \includegraphics[width=0.45\textwidth]{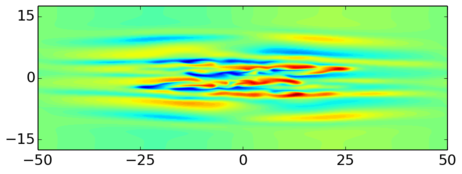} \hspace{2mm}
{\footnotesize (f)}  \hspace{-2mm} \includegraphics[width=0.45\textwidth]{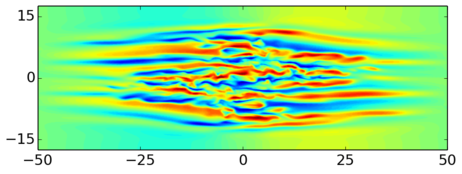} \\
\caption{{\bf Evolution of unstable eigenfunctions.} (a,d) show $\|\bu(t) \|$
versus time $t$ for small perturbations of the doubly-localized solution along its most 
unstable eigenfunctions, at (a) $\Rey = 380$ and (d) $\Rey = 400$. Color denotes 
the symmetries of the eigenfunction perturbations: 
{\color{red} red} for \revision{$\langle \sxy, \sz \rangle$} symmetric eigenfunctions, 
black for \revision{$\langle \sxy, -\sz \rangle$},
{\color{blue} blue} for \revision{$\langle -\sxy, \sz \rangle$}, 
{\color{green} green} for \revision{$\langle -\sxy, -\sz \rangle$},
and {\color{cyan} cyan} for perturbations along 
combinations of eigenfunctions that break all symmetries. Solid lines indicate the 
most unstable eigenfunction of each symmetry group, and dashed are the second-most unstable. 
The midplane streamwise velocity of the most unstable perturbation at (b) $t=0$ and (c) 
$t=200$ for $\Rey=380$, and (e,f) the same for $\Rey=400$.
}
\label{f:instabilities}
\end{figure}

\refFig{f:instabilities} characterizes the temporal evolution 
\revision{$\bu(t) = f^t(\bu_0 + \delta \bu)$ of the doubly-localized solution 
$\bu_0$ perturbed} along its most unstable eigenfunctions and along 
combinations of eigenfunctions that break all symmetries. Perturbation 
magnitudes were set to $\| \delta \bu \| /\| \bu_0 \| = 10^{-2}$, where 
$ \|\bu\|^2 = 1/V \int_V \bu \cdot \bu \; d\bx$, with $V$ the volume of the 
computational domain. For $\Rey \leq 360$ (not shown) most perturbations 
\revision{produce a short period ($t < 200$) of transient growth, but 
in all cases monotonic relaminarization $\| \bu(t) \| \rightarrow 0$
begins by $t \approx 500$. At $\Rey = 380$, several perturbations 
produce long-lived $(t > 1000)$ turbulent spots, and at $\Rey = 400$ all perturbations do.}
\refFig{f:instabilities}(b,c) shows a typical decaying spot at $\Rey = 380$, and 
\reffig{f:instabilities}(e,f) shows a typical growing spot at $\Rey = 400$.

\revision{Unlike edge states, the doubly-localized solution has stable manifold 
of co-dimension greater than 1 for all Reynolds numbers (even with $\langle \sxy, \sz \rangle$ 
symmetry restriction), so the stable manifold cannot divide state space and form a 
laminar/turbulent boundary by itself. However, the fact that some perturbations 
from the doubly-localized solution lead to laminar flow and some to turbulence 
demonstrates that the solution lies on the laminar/turbulent boundary,
and that portions of its unstable eigenspace lie on either side of the boundary,
for the approximate range $360 \lesssim \Rey \lesssim 400$. For $\Rey \leq 360$,
the solution lies on the laminar side of the boundary, and $\Rey \geq 400$,
it lies on the turbulent side.} In all cases the perturbations rapidly generate 
fine-scale structure in the velocity field, which then either decays or 
grows in a complex, long-term, and perturbation- and Reynolds-dependent manner. 
Note that oblique perturbations 
\revision{(with $\sxyz \in \langle -\sxy, -\sz \rangle$ symmetry)}
produce turbulent spots with $\sxyz$ symmetry but little noticeable 
obliqueness (\reffig{f:instabilities}(c,f)), though eventually such spots can
grow to fill the computational domain with a pattern of tilted laminar/turbulent 
bands.

%% file: conclusions.tex
\section{Conclusions}
\label{s:conclusions}

We have computed a doubly-localized solution of plane Couette flow, 
which consists of two symmetrically-opposed $\Lambda$-shaped vortices 
whose legs swirl in opposite directions and are tilted in both the 
spanwise and wall-normal directions. The solution roughly resembles in
size and internal structure the smallest sustained turbulent spots 
simulated by \citet{LundbladhJFM91} at similar Reynolds number and 
the doubly-localized, chaotically wandering edge states of \cite{DuguetPF09} 
and \cite{SchneiderJFM10}. The streamwise exponential decay and the 
form of the wall-normal overhang profile are well-approximated by the
solution of a linearized equation involving the leading terms of the 
Navier-Stokes equations. Over a range of Reynolds numbers the solution 
\revision{lies} on the boundary between states that decay to laminar flow 
and those that grow to turbulence.